\documentclass[twocolumn]{emulateapj}
\usepackage{amsfonts,amsmath,amssymb}
\usepackage{graphicx,dcolumn,bm}
\usepackage{mathptmx}                    

\def\ms{M_\odot}
\def\mtov{M_\text{TOV}}
\def\tl{\tilde\Lambda}

\begin{document}

\title{
Are small radii of compact stars ruled out by
GW170817/AT2017gfo?}

\author{
G. F. Burgio,$^1$ A. Drago,$^2$ G. Pagliara,$^2$ H.-J. Schulze,$^1$
and J.-B. Wei$^1$}

\affiliation{
$^1$ INFN Sezione di Catania, Dipartimento di Fisica,
Universit\'a di Catania, Via Santa Sofia 64, 95123 Catania, Italy,\\
$^2$ Dip.~di Fisica e Scienze della Terra dell'Universit\`a di Ferrara
and INFN Sez.~di Ferrara, Via Saragat 1, I-44100 Ferrara, Italy}

\begin{abstract}
The detection of GW170817 and its electromagnetic counterparts allows
to constrain the equation of state of dense matter in new and
complementary ways. Very stiff equations of state are ruled out by
the upper limit on the average tidal deformability, $\tl\lesssim 800$,
imposed by the detected gravitational wave signal. A lower limit,
$\tl\gtrsim 400$, can also be extracted by considering the large
amount of ejected matter which powers the kilonova AT2017gfo.  By
using several microscopic nucleonic equations of state, we first
confirm the existence of a monotonic relation between $R_{1.5}$ (the
radius of the $1.5\ms$ configuration) and $\tl$.  This translates the
limits on $\tl$ into limits on the radius: $11.8\,\text{km} \lesssim
R_{1.5} \lesssim 13.1\,$km.  We then show that the monotonic relation
is violated, if a second branch of compact stars composed of quark
matter exists, as in the two-families or the twin-stars scenarios.  In
particular, it is possible to fulfill the limits on $\tl$ while having
$R_{1.5}$ significantly smaller than $12\,$km.  In both those
scenarios the event GW170817/AT2017gfo originates from the merger of a
hadronic star and a star containing quark matter.
\end{abstract}

\maketitle

\section{Introduction}

The detection of the first signal of gravitational waves (GWs) from
the merger of two neutron stars (NSs) in August 2017, GW170817
\citep{TheLIGOScientific:2017qsa}, has clearly shown how powerful is
this new observational tool to study the properties of dense matter
and its equation of state (EOS).  Indeed, it was possible to set the
first upper limit on the dimensionless tidal deformability
$\Lambda_{1.4}$ of a NS with a mass of $1.4\ms$: $\Lambda_{1.4}<800$
at $90\%$ confidence level (for the case of low-spin priors).  As a
general rule, stiff EOSs lead to NSs with large radii which are easily
deformed by the tidal field of the companion and have,
correspondingly, a large value of $\Lambda$.  In
\citet{TheLIGOScientific:2017qsa}, it has been shown that some very
stiff EOSs such as MS1 and MS1b \citep{Mueller:1996pm} are basically
ruled out.  A number of analyses have confirmed this conclusion:
\citet{Annala:2017llu}, by using a general polytropic parametrization
of the EOS which is compatible with perturbative QCD at very high
density, have shown that $\Lambda_{1.4}<800$ implies that the radius
of a $1.4\ms$ compact star is $R_{1.4}<13.4\,$km. Similar results
have been obtained by \citet{Most:2018hfd} and by \citet{Lim:2018bkq},
where an EOS based on chiral effective field theory has been used up
to densities close to nuclear matter saturation density. Also
\citet{raithel} confirm these findings.

The source of GW170817 has released also two strong electromagnetic
signals: a short gamma-ray burst GRB170817A delayed by $\sim2\,$s with
respect to the GW signal and a kilonova, AT2017gfo, with a peak of
luminosity occurring a few days after the merger
\citep{GBM:2017lvd,Monitor:2017mdv}. The short gamma-ray burst did
not show any prolonged activity after the prompt emission: the
merger's remnant is therefore most likely a hypermassive star that in
less than one second has collapsed to a black hole
\citep{Margalit:2017dij,Ruiz:2017due,Rezzolla:2017aly}. 

On the other hand,
one can infer that also extremely soft EOSs are ruled out.
A first argument is again based on the observation of the short gamma ray burst:
the post-merger remnant did not collapse promptly to a black hole
but survived as a hypermassive star for at least a few ms.
In turn, this implies that the total mass of the binary system was below
the threshold mass for prompt collapse $M_\text{th}$%
\footnote{\label{note}
Within the two-families scenario
\citep{Drago:2013fsa,Drago:2015cea,Drago:2015dea,Wiktorowicz:2017swq}
this implies that the event of August 2017 was not the merger of two NSs,
but the merger of a hadronic star and of a strange quark star
\citep{Drago:2017bnf}.}.
The value of $M_\text{th}$ depends quite strongly on the adopted EOS:
the softer the EOS the lower the value of $M_\text{th}$.
\citet{Bauswein:2017vtn} found that GW170817 allows to set
a lower limit on the radius of the $1.6\ms$ compact star:
$R_{1.6}>10.7\,$km,
thus excluding very soft EOSs.

Also the observed kilonova signal provides constraints on the EOS.
The kilonova
\citep{Cowperthwaite:2017dyu,Nicholl:2017ahq,Alexander:2017aly,Pian:2017gtc,2017Sci...358.1556C}
is generated by the mass ejected from the merger.
\citet{Radice:2017lry} inferred that the large amount of ejected matter,
needed to explain the features of AT2017gfo,
implies a not too soft EOS.
In particular, the average tidal deformability of the binary,
\begin{equation}
 \tilde \Lambda =
 \frac{16}{13} \frac{(M_1+12M_2)M_1^4}{(M_1+M_2)^5} \Lambda_1
  + (1 \leftrightarrow 2) \:,
\label{e:lambda}
\end{equation}
where $M_1$ and $M_2$ are the masses of the components,
must be larger than about $400$.
By using the results of \citet{Annala:2017llu} one obtains that
$R_{1.4}\gtrsim 12\,$km.

Another way to impose limits on the smallest possible value of $R_{1.4}$
is based on incorporating all lab information about the EOS
at densities up to saturation.
For instance, the PREX collaboration obtained a measurement
of the neutron skin of $^{208}$Pb \citep{Abrahamyan:2012gp},
and from this measurement \citet{Fattoyev:2017jql} derived constraints
on the density dependence of the symmetry energy which, in turn,
translate into limits on the values of the radius and of the tidal deformability:
they obtain $R_{1.4}>12.55\,$km and $\Lambda_{1.4}>490$.
Somehow similarly,
the very recent analyses of \citet{Most:2018hfd} and \citet{Lim:2018bkq}
confirm a lower limit for $R_{1.4}\gtrsim$ (11.65-12)\,km
by using state-of-the-art EOSs at subnuclear densities.
Note that all these limits are obtained by assuming
that only one family of compact stars exists
and that no first-order phase transition to quark matter occurs
at large densities.

A more ``traditional" technique to constrain the radii of compact
stars relies on the modelling of the X-ray spectra of compact stars in
LMXBs.  Some analyses indicate very small radii: for stars of
(1.4-1.5)$\ms$, the review paper of \citet{Ozel:2016oaf} suggest radii
in the range (9.9-11.2)\,km.  Those results have been criticized in
\citet{Steiner:2010fz,Lattimer:2013hma}: in particular, if the
atmosphere contains He, significantly larger radii are extracted
\citep{Lattimer:2013hma}.  More recently, \citet{Steiner:2017vmg} have
shown that when allowing for the occurrence of a first-order phase
transition in dense matter (Model C), $R_{1.4}$ is smaller than
$12$\,km to $95\%$ confidence, confirming a previous analysis of \citet{Steiner:2010fz}.
However, $R_{1.4}$ could be larger if
neutron stars have uneven temperature distributions. Clearly, no firm
conclusions can yet be reached and we need to wait for new data such
as the ones collected by the NICER mission.

In this Letter, we investigate under which conditions
$R_{1.5}<12\,$km can be consistent with the limits on $\tl$
extracted from GW170817/AT2017gfo.
As suggested also by \citet{Fattoyev:2017jql},
the tension between small radii and not too small $\tl$
can be relieved if a strong phase transition occurs at supranuclear densities.

We first present results for the mass-radius relations and
tidal deformabilities of neutron stars as obtained
by microscopic calculations of the EOS.
We show that $R_{1.5}$ is typically larger than about $11.8\,$km
for $\tilde \Lambda \gtrsim 400$
(if the maximum mass is larger than about $2\ms$).
We then explore two possibilities for the appearance of quark matter
in compact stars:
a first scenario based on the coexistence of two families of compact stars,
i.e., hadronic stars (HSs) and quark stars (QSs)
\citep{Drago:2013fsa,Drago:2015cea,Drago:2015dea,Wiktorowicz:2017swq},
and a second scenario based on the so called ``twin-stars" solution
of the Tolman-Oppenheimer-Volkoff (TOV) equation,
obtained in presence of a strong first-order phase transition to quark matter
\citep{Schertler:2000xq,Alford:2015dpa,Paschalidis:2017qmb}.
We discuss how the formation of quark matter
allows to fulfill the constraints on $\tl$
and to obtain at the same time stellar configurations
with radii significantly smaller than $11.5\,$km
in the two scenarios discussed above.

\section{Equations of state of dense matter}

Let us first discuss the standard one-family scenario
and the modeling of the EOS.
At variance with recent calculations using only polytropic EOSs or
only phenomenological EOSs with parameters fitted to
properties of nuclear matter and finite nuclei around saturation density,
we use here also the more reliable `microscopic' EOSs based on
many-body calculations.
In particular we examine several EOSs \citep{Li:2008zzt}
obtained within the Brueckner-Hartree-Fock (BHF)
approach to nuclear matter \citep{1976Jeu,1999Book,2012Rep},
which are based on different nucleon-nucleon potentials
[the Argonne $V_{18}$ (V18,UIX) \citep{Wiringa:1994wb},
the Bonn B (BOB) \citep{bonn1},
and the Nijmegen 93 (N93) \citep{nij1,nij2}]
and compatible three-nucleon forces \citep{glmm,uix,zuo,tbfnij} as input.
Furthermore we compare with the often-used results of the variational
calculation (APR) \citep{1998APR}
and the Dirac-BHF method (DBHF) \citep{dbhf1,dbhf2,dbhf3},
employing V18 and Bonn A potentials, respectively.
Two phenomenological relativistic-mean-field EOS are used for comparison:
LS220 \citep{Lattimer:1991nc} and SFHo \citep{Steiner:2012rk}.

Apart from these purely nucleonic EOSs we also examine EOSs containing
hyperons, BOB(NN+NY) \citep{mmy,chen11} and V18(NN+NY+YY) \citep{mmyy}.
Finally, the SFHo with the inclusion of delta resonances and hyperons (SFHO+HD)
is also analyzed \citep{Drago:2014oja}:
in particular we consider two parametrizations
corresponding to two different values for the coupling of the delta
resonances with the sigma meson:
$x_{\sigma\Delta}=1.15$ (SFHO+HD) and
$x_{\sigma\Delta}=1$ (SFHO+HD2),
while we set the couplings with the omega and the rho meson to
$x_{\omega\Delta}=x_{\rho\Delta}=1$.
These two choices are motivated by several analyses of scattering data
(electron and pion scattering off nuclei),
suggesting a coupling with the sigma meson stronger than the coupling
with the omega meson, see the discussion in \citep{Drago:2014oja}.

Concerning the quark-matter EOS we adopt two models
representative for the two-families and twin-stars scenario, respectively:

i) A simple parametrization of a strange-quark-matter EOS (SQM)
encoding both the non-perturbative phenomenon of confinement and the
perturbative quark interactions \citep{Weissenborn:2011qu}.
We consider two parameters sets:
the set QS with $B_\text{eff}^{1/4}=137.5\,$MeV and $a_4=0.7$
whose corresponding maximum mass is $\mtov=2.1\ms$,
and the set QS2 with $B_\text{eff}^{1/4}=142\,$MeV and $a_4=0.9$
whose corresponding maximum mass is $\mtov=2.0\ms$.
For the two-families scenario,
these EOSs are combined with the hadronic SFHO+HD and SFHO+HD2 EOSs
for the low-mass, small-radius partner.

ii) A constant-speed-of-sound EOS (DBHF+CS) to model hybrid stars:
one adopts a nucleonic EOS up to a transition pressure $p_\text{trans}$ and
implements a first-order phase transition,
which is characterized by a energy density jump $\Delta e$
at the onset of the phase transition
and by the speed of sound of pure quark matter $c^2_q$.
For this study we have taken the results of \citep{Alford:2015dpa}
for the DBHF nucleonic EOS
and we have set $p_\text{trans}/e_\text{trans}=0.1$,
$\Delta e/e_\text{trans}=1$ and $c^2_q=1$.
One needs to choose a speed of sound saturating the causal limit,
because with more ``normal" values
it is impossible to obtain $\mtov \geq 2\ms$ and $R_{1.4}\leq 12\,$km.
Still a strong fine-tuning of the parameters is needed
in order to satisfy all constraints.
For comparison we also consider the parameter set DBHF+CS2:
$p_\text{trans}/e_\text{trans}=0.095$,
$\Delta e/e_\text{trans}=0.65$ and $c^2_q=2/3$,
which leads to a larger value of $R_{1.4}$.

\begin{figure}[t]
\vspace{-8mm}\centerline{\hskip5mm\includegraphics[scale=0.36]{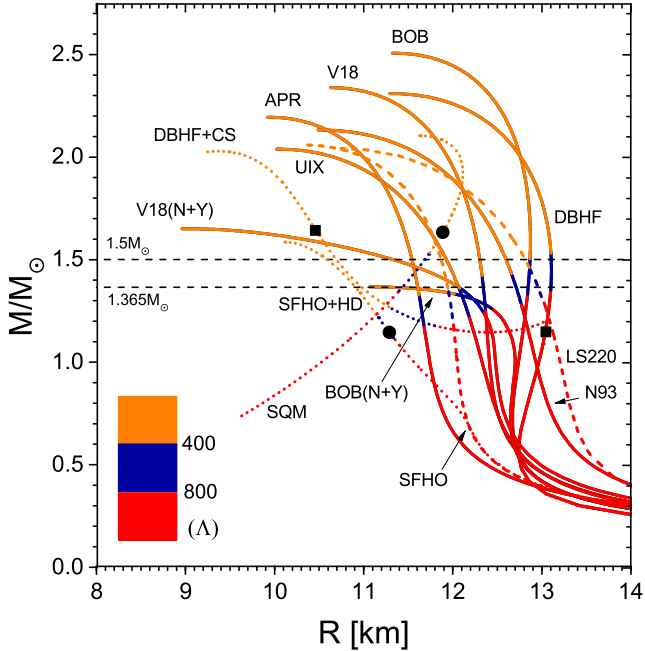}}
\vspace{-5mm}
\caption{(Color online)
Mass-radius relations for different EOS,
indicating also values of the tidal deformability $\Lambda$.
Solid (dashed) curves are for microscopic (phenomenological) EOSs, see text.
Markers indicate the $q=0.7$ configurations for the
two-families ({$\bullet$}) and twin-stars ({\tiny$\blacksquare$}) scenarios.
}
\label{f:lr}
\end{figure}

\begin{figure}[t]
\vspace{-9mm}\centerline{\hskip8mm\includegraphics[scale=0.36]{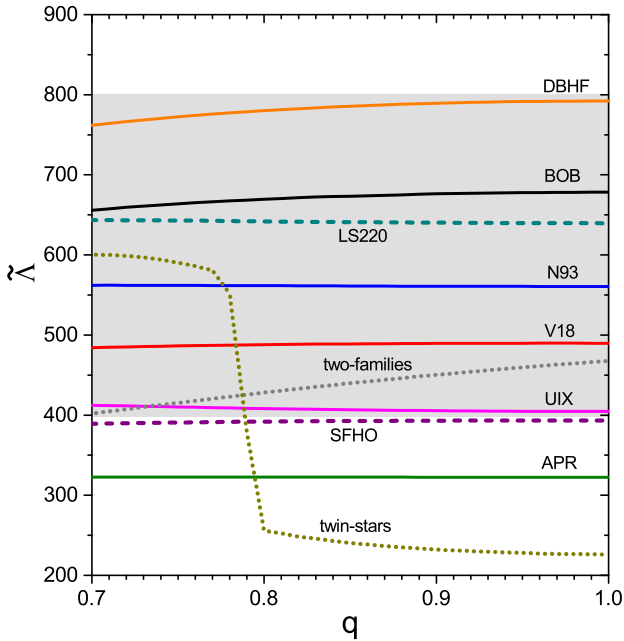}}
\vspace{-9mm}
\caption{(Color online)
Effective deformability $\tl$,
Eq.~(\ref{e:lambda}),
vs.~mass asymmetry $q=M_2/M_1$
for a binary NS system with fixed chirp mass $M_c=1.188\ms$
for different EOSs.
The shaded area is constrained by the interpretation of the GW170817 event.
}
\label{f:lq}
\end{figure}

\begin{figure}[t]
\vspace{-7mm}\centerline{\includegraphics[scale=0.36,clip]{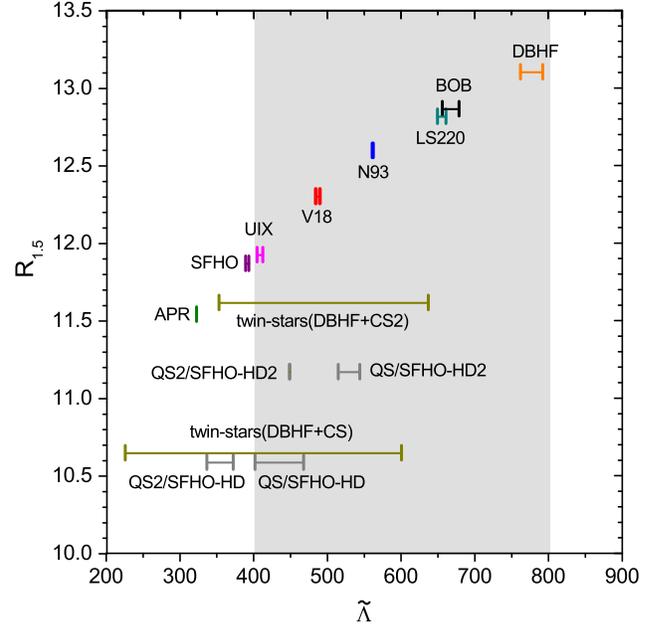}}
\vspace{-7mm}
\caption{(Color online)
Smallest $R_{1.5}$ radius of an asymmetric binary NS system
and possible range of $\tl$
with fixed chirp mass $M_c=1.188\ms$ and varying $q=0.7-1$
for different EOSs.
For the two-families scenario we show the results obtained
when combining SFHO-HD or SFHO-HD2 with QS or QS2 (in the case QS2/SFHO-HD2 there is no visible dependence on q).
}
\end{figure}

\section{Results and discussion}

For each EOS we construct the family of solutions of the TOV system
with the addition of the equation for the tidal Love number $k_2$
\citep{Hinderer:2009ca,Postnikov:2010yn},
which is related to the dimensionless tidal deformability by
$\Lambda\equiv(2/3)(R/M)^5 k_2$.

In Fig.~1 we display the mass-radius relations for the EOSs here adopted
and we encode also the information on the tidal deformabilities.
Note most importantly
that EOSs which reach the two-solar-mass lower limit
and fulfill the constraint $400<\Lambda_{1.365}<800$,
predict $12\,\text{km} \lesssim R_{1.5} \lesssim 13\,$km,
in agreement with the analysis of \citet{Annala:2017llu}.

Note also that not all the EOSs satisfy the two-solar-mass limit:
some EOSs in which also hyperons and/or delta resonances are included
(in particular SFHO-HD) lead to small maximum masses
($\mtov\approx 1.6\ms$) and, at the same time,
to very compact configurations, $R_{1.5}<11\,$km.
Such EOSs would be excluded within the standard one-family scenario in which
all compact stars belong to the same family:
in that scenario there is a one-to-one correspondence
between the mass-radius relation and the EOS.
However, they are allowed if one adopts the so-called two-families scenario
in which the heaviest stars are interpreted as QSs,
whereas the lighter and smaller stars are interpreted as HSs
\citep{Drago:2013fsa,Drago:2015cea,Drago:2015dea,Wiktorowicz:2017swq}.

To constrain the EOSs by using the data of the event GW170817,
we fix now the chirp mass
$M_c \equiv (M_1 M_2)^{3/5}/(M_1+M_2)^{1/5} = 1.188\ms$
(corresponding to $M_1=M_2=1.365\,\ms$ for a symmetric binary system),
and compute $\tl$, Eq.~(\ref{e:lambda}),
as a function of the mass asymmetry $q=M_2/M_1$.
The range deduced for GW170817 was $q=0.7-1$ \citep{TheLIGOScientific:2017qsa},
corresponding to a maximum asymmetry $(M_1,M_2)=(1.64,1.15)\ms$.
The results are displayed in Fig.~2.
The one-family EOSs predict an effective deformability nearly
independent of the asymmetry $q$.
When calculating $\tl$ within the two-families scenario,
we assume that the binary system is a mixed system
with a light HS and a heavy QS, i.e., $M_1$ refers to a QS (see footnote \ref{note}).
Similarly, within the twin-stars scenario we assume that the most massive
star is the hybrid star,
as in \citep{Paschalidis:2017qmb}.
The configurations corresponding to maximum asymmetry $q=0.7$
are indicated by markers in Fig.~1 for both scenarios.
In particular the twin-stars configuration features a very large difference
between the radii of the two components:
$(R_1,R_2)=(10.7,13.0)\,$km,
which allows to achieve concurrently a very small radius $R_1$
and a sufficiently large $\tl\approx600$.
Notice that in both these scenarios a not negligible dependence
of $\tl$ on $q$ is found.

Fig.~3 summarizes the main outcome of this study:
we display for the different EOSs the correlation between the values of $\tl$
(for a fixed $M_c=1.188\ms$ and with $q$ varying in the range 0.7-1)
and $R^{}_{1.5}$.
In the case of the two-families or twin-stars scenarios,
$R^{}_{1.5}$ indicates the radius of the most compact component.
Within the one-family scenario,
one observes a very tight and monotonic correlation between $R_{1.5}$ and $\tl$.
All the EOSs which fulfill the constraint $\tl>400$
from \citet{Radice:2017lry} lead to $R_{1.5}>11.8\,$km.
This feature is violated if a second branch of compact stars exists,
as in the case of the two families or of the twin stars.
Moreover, for some choice of the parameters,
it is possible to satisfy $\tl>400$ and obtain stellar configurations
with $R_{1.5}$ significantly smaller than $12\,$km.

Within the twin-stars scenario, an extremely detailed parametric
analysis was already performed in \citet{Alford:2015dpa} by using the
nucleonic EOSs DBHF and BHF \citep{Taranto:2013gya}. There, it was
stressed that to obtain $R_{1.5}$ smaller than 12km, $c^2_q$ must be
significantly larger than $1/3$. In Fig.3, we have implemented an
example (DBHF+CS2) for which $R_{1.5}= 11.6$km, obtained by fixing
$c^2_q=2/3$. To reach smaller values of $R_{1.5}$, even larger sound
speeds should be assumed. In the causal limit, $c^2_q=1$, one obtains
$R_{1.5}=10.7$km. We have considered here only the nucleonic EOS DBHF
because BHF is rather a soft EOS and it would not be possible to
satisfy the limit $\tl>400$. In conclusion, the twin-stars scenario
allows to reach radii smaller than $12$km, while satisfying the limit on $\tl$,
only for a very small parameter space.

For the two-families scenario, conversely, the parameter space
is larger. We can fulfill the limit $\tl>400$ with both the hadronic
EOSs SFHO+HD and SFHO+HD2 which lead to $R_{1.5}=10.6 $km and to $R_{1.5}=11.2$km, 
respectively. Only when combining the soft hadronic EOS SFHO+HD
with the soft quark EOS QS2, the limit on $\tl$ is not satisfied.
Notice that in both quark EOSs $c^2_q \sim 1/3$.

Let us now compare the two-families and the twin-stars scenarios.
In the two families the low-mass objects are made of hadrons and the
presence of delta resonances and/or hyperons allows to reach small radii
(and very small values of $\Lambda$)
for masses in the range (1.4-1.5)$\ms$.
The more massive stars are instead QSs and their radii
are not extremely small
(their $\Lambda$ has an intermediate value).
In the twin-stars scenario the low-mass objects are made of nucleons and
have large radii and large $\Lambda$,
while the most massive stars are hybrid stars
with a very large quark content and small radii and $\Lambda$.
Note how in both these scenarios the event of August 2017
needs to be interpreted as a ``mixed case,"
in which one of the objects is made only of
hadrons and the other contains deconfined quarks.
While these two scenarios are both able to interpret the event of August 2017
and to have very small values for $R_{1.5}$,
the differences in their mass-radius relation and in their composition
will provide different and testable outcomes for the three cases of mergers
they are able to produce:
HS-HS, HS-QS, QS-QS in the case of the two-families and
NS-NS, NS-hybrid star, hybrid star-hybrid star
in the case of the twin-stars
(see \citet{Drago:2018nzf} and work in preparation).
For instance, in the case of a merger of two light compact stars, e.g. $1.2\ms+1.2\ms$,
the twin-stars scenario predicts very large values of $\tl$ while
for the two-families scenario $\tl$ is significantly smaller. This difference can easily be tested
both through the GW signal and through the kilonova.

An open issue regarding both the two-families and the twin-stars
scenario concerns the estimates of the mass ejected during/after the
merger and its correlation with $\tl$, in the case of a mixed system,
HS-QS or NS-hybrid star. Presently, no numerical simulation has been
performed for these two scenarios. In particular, within the
two-families scenario the numerical task is very challenging because
one has to deal with two different EOSs. However, in both cases
GW170817 can be interpreted as due to the merger of quite an
asymmetric system, $q\lesssim 0.8$, in which the low-mass component
has a value of $\Lambda$ exceeding $400$. Asymmetric systems lead in
general to a larger amount of ejected matter with respect to symmetric
systems \citep{Bauswein:2013yna}. Moreover, since the low-mass
companion is a HS with a not too small $\Lambda$, we expect that most
of the ejected mass is provided by the tidal disruption of the
hadronic star and that the accretion torus which forms around the
post-merger remnant is made of hadronic matter. It will be
interesting, in future calculations, to investigate whether the event
AT2017gfo can be explained through the material ejected from the HS
component of the mixed system in the two-families scenario.

\section{Conclusions}

While the standard interpretation of the GW170817 event in the
one-family scenario is perfectly compatible
with the merging of two nucleonic neutron stars
governed by a microscopic nuclear EOS respecting the $\mtov>2\ms$ limit,
we have shown here that the lower limit on the tidal deformability obtained
by \citet{Radice:2017lry} is not incompatible with $R_{1.5}$ even significantly smaller than 12\,km
if one assumes that the population of compact stars is not made of only one
family.
Indeed, when allowing for the existence of disconnected
branches in the mass-radius relation,
either within the two-families scenario or within the twin-stars scenario,
one can explain the existence of very compact stars and at the same time
one can fulfill the request of having a not too small
average tidal deformability, as suggested by the analysis of AT2017gfo.

Notice that in both scenarios,
the source of GW170817 is a mixed binary system:
a hadronic star and a quark star within the two-families scenario
\citep{Drago:2017bnf} and a hybrid star and a nucleonic star within
the twin-stars scenario \citep{Paschalidis:2017qmb}.
It is interesting to note that within the two-families scenario,
a system with the chirp mass of the source of GW170817 cannot be composed of
two hadronic stars:
such a system would have a too small average tidal deformability
and moreover it would lead to a prompt collapse \citep{Drago:2017bnf}.
In this respect, the constraint on the tidal deformability
and the evidence of formation of a hypermassive star
within the event GW170817 are both suggesting that one of the two
stars must be a quark star,
if the hypothesis of the two families of compact stars is adopted.

Most of the analyses suggesting limits on the radii are
based on a statistical average of a few stellar objects.
It is therefore possible that some of those objects have radii
even significantly larger than the average.
A feature of the two-families scenario concerns the mass
distribution of hadronic stars and quark stars.
It exists a ``coexistence mass range''
in which compact stars can be both hadronic stars or quark stars.
In \citet{Wiktorowicz:2017swq} a population-synthesis analysis has shown
that the fraction of quark stars in LMXBs in that mass range is not marginal,
up to 30\% or more. It is therefore possible that some of the ``neutron stars" in a LMXB
are actually quark stars and have a
radius of about (11.5-12)\,km.
The analysis of the X-ray signal emitted by a quark star in a LMXB
is though still a rather unexplored problem \citep{Zdunik:2001yz,Kovacs:2009gt}
and it is at the moment very difficult to indicate how to distinguish
a quark star from a neutron star only from the properties of its X-ray spectrum.
It is interesting to remark that \citet{Nattila:2017wtj} suggest for the neutron star in 4U 1702−429
a radius of (12.4$\pm$0.4)km for a mass of (1.9$\pm$0.3$)\ms$, nicely sitting on our
quark star branch.

Future measurements of GWs from binary collisions could give us
even tighter constraints on the tidal deformability by accumulating data
from several GW events \citep{Agathos:2015uaa}.
Moreover, very soon, also the NICER collaboration will release results for
the measurements of the radii of the closest pulsars.
In the near future we expect therefore to have
a crucial opportunity to test the hypothesis
that quark matter does form in compact stars.
Finally, another complementary way to study the properties of dense matter
relies on the measurement of the moment of inertia:
the future SKA experiment will face this task.
We will present in a forthcoming paper our predictions
for the moment of inertia of compact stars in the standard
one-family scenario and in the two-families scenario.


\end{document}